# MODetector (MOD): A Dual-Functional Optical Modulator-Detector for On-Chip Communication

**Shuai Sun[1], Ruoyu Zhang[1], Jiaxin Peng[1], Vikram Narayana[1], Hamed Dalir[2], Tarek El-Ghazawi[1], and Volker J. Sorger[1]**

[1]*Department of Electrical and Computer Engineering, George Washington University, Washington, DC 20052 USA*
[2]*Omega Optics, Inc., 8500 Shoal Creek Blvd., Bldg. 4, Suite 200, Austin, TX 78757, USA*
*sorger@gwu.edu*

**Abstract:** Physical challenges at the device and interconnect level limit both network and computing energy efficiency. While photonics is being considered to address interconnect bottlenecks, optical routing is still limited by electronic circuitry, requiring substantial overhead for optical-electrical-optical conversion. Here we show a novel design of an integrated broadband photonic-plasmonic hybrid device termed MODetector featuring dual light modulation and detection function to act as an optical transceiver in the photonic network-on-chip. With over 10 dB extinction ratio and 0.8 dB insertion loss at the modulation state, this MODetector provides 0.7 W/A responsivity in the detection state with 36 ps response time. This multi-functional device: (i) eliminates OEO conversion, (ii) reduces optical losses from photodetectors when not needed, and (iii) enables cognitive routing strategies for network-on-chips.

## I. INTRODUCTION

The semiconductor industry has pushed the physical limits of a single electronic logic device down to nanometer scale for higher speed and energy efficiency, yet physical limitations remain such as quantum tunneling, current leakage [1-2]. Although new emerging options including photonics and plasmonics have recently been investigated to overcome interconnect bottleneck, the fundamental energy efficiency of optical logic devices is still been limited to fJ level due to weak non-linearity [3]. Rather than focusing at the logic level, reducing the energy cost in data communication can be a more efficient option [4-5]. While nanophotonic and plasmonic building blocks show performance and integration density promise, metal-optics is unlikely to become a mean for implementing optical communication links due to losses. However, separating the function of communication and light manipulation to passive photonics such as Silicon photonics, and plasmonic, respectively, shows superior intra-chip capabilities in terms latency, energy efficiency and packing density [6]. Such hetero-integration shows up to 300% energy improvement and an order of magnitude footprint reduction over traditional photonic networks [7-9]. However, despite these promises, the point-to-point communication characteristic of such photonic heterogeneous integration schemes can be restrictive in realizing on-chip interconnect networks for future many-core chips with hundreds of cores. Conventional photonics, on the other hand, can achieve the required connectivity through a bus structure, relying on wavelength division multiplexing (WDM) to support multiple links on such a shared physical communication medium [10]. However, this is achieved at the cost of large area overheads for microring resonators (MRRs) that aid modulation and detection, as well as substantially higher power consumption attributed to the thermal trimming of the MRRs [7]. Hybrid Photonic-Plasmonic Interconnect (HyPPI) based on our previous demonstration, on the other hand, uses photonic plasmonic hybrid modulators directly integrated onto the photonic transmission waveguide [6]. Although this is beneficial in saving footprint and enhancing point-to-point performance, the on-bus modulator introduces superfluous insertion losses in the level of 1dB/μm even while inactive, rendering the bus unusable to other communication demands. An off-bus modulator is therefore essential. Similarly, on the receiver side, an off-bus detector like those based on photonic microring resonators (MRR) is desirable since it can be actively selected to either coupling to the bus or not, yet associates with tremendous energy and latency overheads due to the electrical thermal tuning. The alternative way of achieving point-to-point connectivity is to connect each core to all the other cores on the chip directly which is however not practical; for instance, a 64-core processor requires a total of 4,032 unidirectional HyPPI links to achieve full connectivity which is several orders of magnitude footprint consuming than the mesh topology [11-12].

To overcome the limitation of point-to-point connectivity while retaining high performance of all-photonic routing, modulation, and detection, we here show a novel photonic-plasmonic hybrid device termed MODetector (hereinafter referred to as 'MOD') that combines both the modulation and the detection functions for inter-chip communication. Different from the regular photonic or plasmonic enhanced (e.g. HyPPI) interconnect, where the modulators and detectors are integrated onto the waveguide bus, this MOD-based HyPPI link allows data to bypass photonic devices without causing much energy penalties [6]. Thus, from a network point of view, we re-allocate some of the routings onto the device level by combining the electro-optic modulation, detection, and optical routing into one single structure. The remainder of the paper is organized as follows: 1) Network comparison between classic point-to-point (P2P) optical links and MOD-enabled links. 2) MOD structure introduction. 3) Operation principle discussion. 4) Modulation and detection analysis.

## II. Towards 'Off-bus Communication'

Improving the links energy efficiency has become a more effective and reliable way of improving the power efficiency on-chip, than reducing the energy consumption of the logic gates due to the inherent static power dissipation arising from quantum tunneling across thin oxides [5]. In recent optical network on chip (NoC) studies, routing protocols based on circuit switching is used as the 'express lines' for high speed, long distance communication [7-11]. Here the actual setup of the physical light path is for circuit switching done by spectrally sensitive devices such as MRRs or hybrid photonic-plasmonic based devices [13-14]. Note, it is worth to mention that, for certain types of the network topologies such as the mesh and the ring, the ability to realize an all-to-all core communication in a power and latency efficient way is highly desirable, which is also been considered as the design goal of our device.

Although traditional ring-based optical NoCs could achieve any core to any core connections with the rings sitting on the joint bus thermally tuned into the same resonance, this requires a long setup time up to millisecond level that depending on the thermal tuning power and the quality (Q) factor of the ring with an average energy consumption as high as 40% of the entire interconnect energy consumption [12-13]. Moreover, for an optical network in general, the all-to-all connection can only be achieved by 1) building a mesh network with 18 links that connect to all the adjacent cores or 2) connecting each core with all the other cores with 72 P2P links (Fig. 1a and Fig. 1b). Note, since the P2P link is unidirectional, every two cores require 2 links to communicate with each other (i.e. simultaneous communication demanded). The first option (regular mesh Fig. 1a) reduces the number of links needed by detecting and regenerating the signal at each core, while the second option (all-to-all mesh, Fig. 1b) saves the OEO conversions by making all the connections available at the same time. This would however not only require larger on-chip footprint but more importantly need massive routers; for our small all-to-all example network one would require a 9×9 router, which is significantly more complex than regular size routers (e.g. 5×5) used regularly in optical networks and scales with a factor of $N(N-2)$ where $N$ is the number of ports needed for the router [Lin12]. Besides, all the OEO conversions in the cores of the P2P-based mesh network and the local routing mechanism in the P2P-based all-to-all network introduce on average 12 dB and 5 dB optical loss per core respectively with high-efficiency hybrid photonic-plasmonic integrated technology [15]. Thus, the P2P interconnect option with both modulation and detection devices on the bus does not allow for reduced network complexity, and hence limits scaling and is not suitable for designing high-connectivity NoCs.

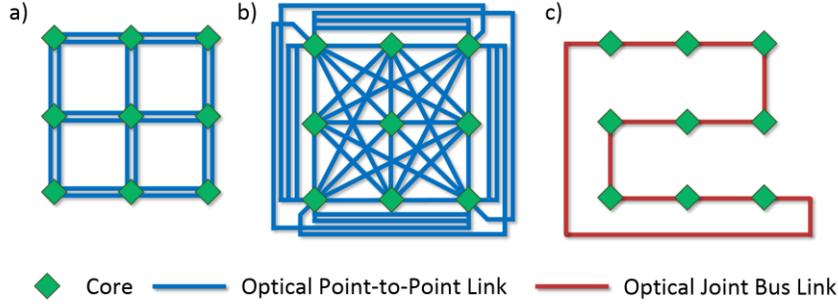

Fig. 1. The optical 3×3 network topology comparison among: a) the P2P-based mesh network; b) the P2P-based all-to-all network and c) the MOD enabled ring network.

Given the aforementioned drawbacks, an off-bus device that can be actively bypassed when its functionality is not needed as discussed in this paper. For the same 3×3 network, only a single link is used that creates a joint bus to connect all 9 cores (Fig. 1c). This off-bus device significantly simplifies the network design while maintaining every core to every core connectivity. We find that the required overhead for this network is as low as insertion loss of a single 2x2 switch per core when the signal needs to be bypassed. Furthermore, this reconfigurable MOD device can be a supplement for the Dynamic Data Driven Applications Systems (DDDAS) enabling configuration based on executing applications and real-time traffic feedback controls [16]. Since MOD is plasmonic and hence spectrally broadband, the joint bus could support multiple wavelengths (i.e. WDM) with even higher connectivity and bi-section bandwidth that communicate simultaneously if the MOD connects to a wavelength selective device (e.g. micro-ring resonator).

### III. On-chip Optical Transceiver Design

#### A. MOD Structure

The goal of MOD is to separate the light modulation and detection from the main bus in order to avoid the unnecessary conversions between electrical and optical domains leading to losses. Since there is no conflict in separating the modulator from the bus and the detector from the bus, it is beneficial to integrate both functionalities into a single device. Moreover, for network topologies like mesh, ring, and bus, some of the cores need to send and receive signals from both directions of the bus, which requires the MOD design to be symmetric in order to enable bi-directional communication. Based on these requirements, we consider a racetrack ring-based MOD structure that integrates an 'expanded' germanium photodetector on the ring via a 2×2 hybrid plasmonic 3-waveguide switch to provide modulation functionality (Fig. 2a) [17]. The 2×2 switch consists of a central switching island containing a highly optical index changeable material (indium tin oxide (ITO)) 'sandwiched' between two gate oxide layers ($SiO_2$) to form a metal-oxide-ITO-oxide-semiconductor capacitive heterostructure; whereas the detector has a germanium block on top of the racetrack waveguide with that part of the silicon etched down to 100 nm for better light mode overlap with the high absorption region (Fig. 2b).

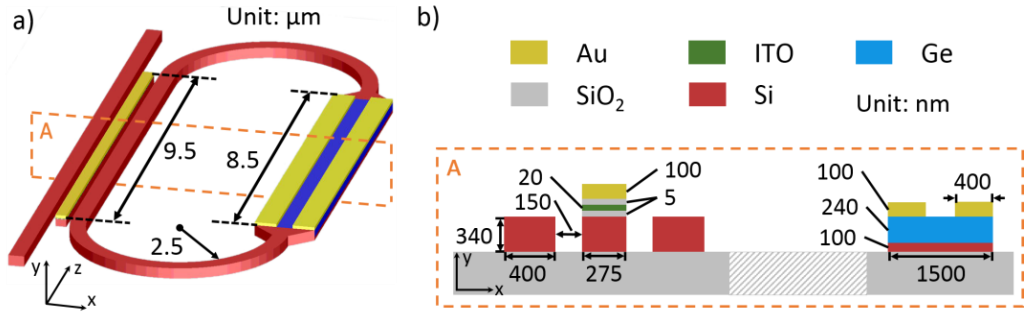

Fig. 2. Schematic plot of MO-Detector. a) The 3D overview of MOD with the ITO hybrid switch on the left and Ge photodetector on the right. b) The cross-section of MOD at $A$ plane. Both a) and b) are color-coded and sharing the same legend on the top-right. All the parameters are optimized for the highest coupling efficiency [17].

### B. Operating Principles

In order to obtain bi-functionality of an optical transceiver, a bias voltage needs to be applied to MOD at different metal gates (i.e. the switch and the detector); the switch changes from the Cross state with 0V bias voltage which allows the light injected from the bus to be coupled to the racetrack ring to the Bar state that keeps the light in the bus waveguide (i.e. bypassing the MOD). For the Bar state, the refractive index of ITO layer shifts from its dielectric region to the metallic region (i.e. epsilon-near-zero, ENZ) due to the carrier concentration change, and thus yields a new coupling length that prohibits the light transverse through the 3-waveguide switch [17-18]. However, this switch could not enable modulating by itself, since the racetrack ring couples any non-detected photons back to the bus. Note, the ring must be connected back to the bus since MOD is required to enable bi-directional communication. Therefore, the detector also needs to be activated to prevent the light from coupling back when modulating a '1' signal is needed. Similarly, switch and detector should cooperate with each other for other functionalities listed in Fig. 3c-3f.

## IV. MOD Function Analysis

### A. Light Modulation

The key part that determines the overall performance of MOD is the coupling efficiency of the 2×2 switch. With a higher switch efficiency, the Ge detector generates more current with the same amount of optical signal power. In addition, a higher extinction ratio at the bus port could be obtained as well. The design of MOD to deploy hybrid photon-plasmon modes is driven by two main factors; first, its compact structure greatly enhanced the light-matter-interaction (LMI) and therefore the device length is much shorter than conventional photonic switch, which is usually in the range of few hundreds of micrometers [19]. Secondly, it combines both sub-λ confinement and long silicon-on-insulator (SOI) waveguide propagation length, which reduces the insertion loss compared to all-plasmonic designs [20]. The fundamental principle of operating this switch is to use bias voltages (i.e. ITO carrier concentration) to control the coupling length (CL) for the different states (Fig. 2b, 2c). The effective mode indices of the first three fundamental TM supermodes ($TM_1$, $TM_2$, $TM_3$) of the switch are used to determine the actual coupling length and the coupling efficiency based on Eqn. 1 and 2 where light source λ is 1550 nm in this case. Moreover, we sweep the ITO carrier concentration from $10^{19}$ to $10^{21}$ cm$^{-3}$ where $10^{19}$ cm$^{-3}$ can be regarded as zero bias voltage point (Fig. 2d). With a higher voltage applied, the effective index difference between $TM_1$ and $TM_3$ becomes smaller resulting in a CL longer than the actual physical length of the switch, which matches with our operation principle discussion before. The 4V bias voltage (carrier concentration of $6.8\times10^{20}$ cm$^{-3}$) was experimentally obtained in previous work and it is used here in all Cross state simulations [18, 20]. Based on the capacitance of the switch, 24 fJ energy is needed to charge the ITO layer from Cross state to Bar state exceeds 100 GHz theoretical switching speed calculated by the RC delay of the device (where the resistance $R_c$ =500 Ω is assumed).

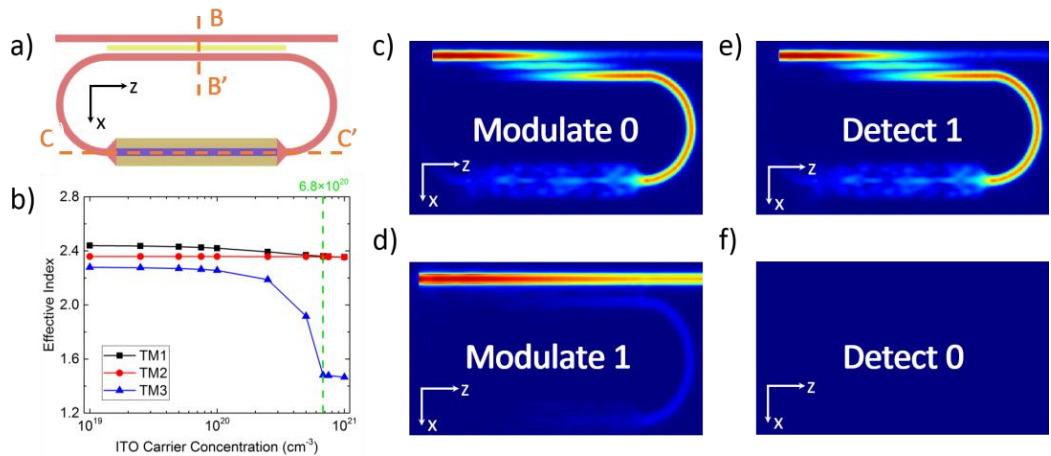

Fig. 3. The switch analysis at the Cross and the Bar states. a) The top view of the MOD with the same color coding as Fig. 1. b) Fundamental TM mode effective indices change of the 3-waveguide switch at the cross-section (BB') based on ITO carrier concentrations. c)-f) The FDTD simulations of all four functionalities at different switch and detector state combinations: c) switch OFF, detector ON; d) switch ON, detector OFF; e) switch OFF, detector ON; e) switch ON, detector ON. All simulations are based on 1550 nm light source. The ITO refractive indices are calculated based on the Drude model. $V_{bias}=V_{dd}=4V$.

$$(n_{TM_1} + n_{TM_2})/2 = n_{TM_3} \tag{1}$$

$$CL = \frac{\lambda}{2(n_{TM_1} - n_{TM_3})} \tag{2}$$

*B. Light Detection*

As for the other main part of MOD, the Germanium (Ge) p-i-n based detector as to functions as well; a) it is a regular photodetector, and b) it is a 'light absorber' to prevent the light from coupling back to the bus waveguide when MOD is used in the modulating mode. In order to achieve high responsivity with CMOS compatible material, we select a metal-semiconductor-metal (MSM) design with Germanium integrated on the SOI waveguide. The choice for Germanium is natural given its high absorption (i.e. high κ) and compatibility with silicon photonics harassing the mature fabrication process [22]. Germanium parameters used are based on the high-temperature-growth process resulting in absorption coefficient ~5000 cm$^{-1}$ [23]. However, based on the TM mode we used for the 2×2 switch, the responsivity is still low if the Germanium is simply placed on the waveguide due to the small mode overlap. Thus, we etch the silicon waveguide down to be just 100 nm thick in the straight part of the racetrack (in red, Fig. 2) and then backfill it with the 240 nm Germanium (Fig. 4a).

Two gold contacts are placed on both sides of the waveguide along the light propagation direction each with 400nm width aligned to the outer side of the waveguide. We noticed that the wider the metal contact (with same doping areas in the Germanium region) results in a faster response speed while at the same time absorbing more light that supposed to be absorbed by the Germanium region thus reduces the responsivity of the detector. To overcome this problem, we expand the Germanium detection region to 1500 nm wide to provide a higher ratio between the light absorbed by Germanium and the light absorbed by gold. Thus, addressing the trade-off between the operating speed and the responsivity by adjusting the metal contact width (Fig. 4b). Nevertheless, the 100 nm contact width provides the highest responsivity, the speed of the detector is only about 20 GHz, which is limited by the longer electron-hole pair moving distance which is determined by the gap between the metal contacts. In addition, not only the contact width but also the carrier generation rate profile affects the speed since electron-hole pairs generated at different positions inside the cross-section of the Si-Ge waveguide varies the carrier-to-contact travel distance. For example, carriers generated from the center of the Germanium have a relatively short distance to the contact, as compared to carriers generated near the contact (insets of Fig. 4b). Thus, a 400 nm contact width with centered carriers achieves our optimized operating speed of 28 GHz with relatively high responsivity and reasonable light leakage and is selected for our detector design. Note, we take the detector speed as the reciprocal of the response time from 10% to 90% of the saturated photocurrent. Although a longer detection region could provide higher responsivity with less light leakage, we aim for a detector length (8.5 μm) similar to that of the 2x2 switch to avoid any additionally required footprint. As mentioned above, the second role of detector region is to prevent the return light leakage to the bus waveguide. Although this detector could not absorb all light for a 400 nm contact width, the 2×2 switch also creates 1.2 dB insertion loss per coupling. Thereby, only 2.8% of the light will leak to the bus at the Cross state, together with another 5% light leakage caused by the unperfected switching, yields over 10 dB extinction ratio together with 83% light bypassing the MOD at the Bar state.

Regarding detector performance, the photocurrent under 0.5V reverse bias is about 0.35 mA under 0.5 mW input light power giving a 0.7 A/W responsivity equivalently while the dark current kept in the sub-μA level (Fig. 4c). Indeed improving the responsivity can be done in multiple ways to include 1) reducing the interface between germanium and gold such as cylindrical metal contacts; 2) using wider rib waveguide, or 3) increasing the absorption of germanium by heating [24-26]. Comparing our optimized detector performance with commercial foundry components, we find a similar matching performance with, for instance, IMECs medium-speed photodetectors process [26]. In addition, the output current can always be amplified by a trans-impedance amplifier to match current requirement for the stage in the next circuit.

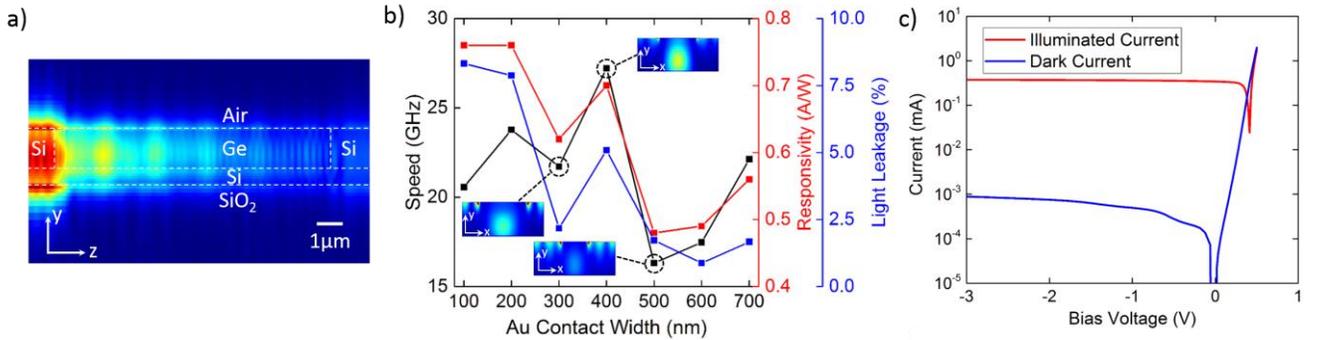

Fig. 4. Detector performance analysis of MOD. a) The cross section FDTD simulated electrical field of the detector part (CC' in Fig. 2a) where light enters from the left side and propagates to the right side. b) The trade-offs between the metal contacts width and the speed, responsivity as well as the light leakages after the detection region. The insets are the generation rate at the same CC' cross section. c) The illuminated and the dark current of the detector at different bias voltages.

## V. Conclusion

By integrating a hybrid photonic-plasmonic switch with a Germanium-based photodetector into one single device, we design a dual-function modulator-detector. This integrated device is able to detect optical signals up to 28 GHz and generate over 100 GHz signals. Based on the symmetric design, it enables bi-directional all-to-all communication between multiple communication cores with only one bus waveguide, which significantly reduces the area for inter-chip connections. With over 10 dB modulation extinction ratio and 0.7 A/W responsivity. This multi-functional device acts an optical transceiver capable of both sending and receiving optical data signals in optical networks and communication links or could potentially be used as a reconfigurable optical element in analog photonic-optical compute engines and accelerators.

## Acknowledgements

This work is supported by the Air Force Office of Scientific Research (AFOSR) award number FA9550-15-1-0447 which is part of the Dynamic Data-Driven Applications System (DDDAS) program and the AFOSR award number FA9550-17-P-0014 of the small business innovation research (SBIR) program.